# BotMosaic: Collaborative Network Watermark for Botnet Detection

Amir Houmansadr[1,1,], Nikita Borisov[1]

*Electrical and Computer Engineering Department*
*University of Illinois at Urbana-Champaign*
*Urbana, IL, 61820, U.S.A*

## Abstract

Recent research has made great strides in the field of detecting botnets. However, botnets of all kinds continue to plague the Internet, as many ISPs and organizations do not deploy these techniques. We aim to mitigate this state by creating a very low-cost method of detecting infected bot host. Our approach is to leverage the botnet detection work carried out by some organizations to easily locate collaborating bots elsewhere.

We created BotMosaic as a countermeasure to IRC-based botnets. BotMosaic relies on captured bot instances controlled by a watermarker, who inserts a particular pattern into their network traffic. This pattern can then be detected at a very low cost by client organizations and the watermark can be tuned to provide acceptable false-positive rates. A novel feature of the watermark is that it is inserted collaboratively into the flows of multiple captured bots at once, in order to ensure the signal is strong enough to be detected. BotMosaic can also be used to detect stepping stones and to help trace back to the botmaster. It is content agnostic and can operate on encrypted traffic. We evaluate BotMosaic using simulations and a testbed deployment.

*Email addresses:* `ahouman2@illinois.edu` (Amir Houmansadr), `nikita@illinois.edu` (Nikita Borisov)

## 1. Introduction

A *botnet* is a network of compromised machines, *bots*, that is controlled by one or more *botmasters* to perform coordinated malicious activity. Botnets are among the most serious threats in cyberspace due to their large size (Ramachandran and Feamster, 2006). This enables the bots to carry out various attacks, such as distributed denial of service, spam, and identity theft, on a massive scale.

Botnets are controlled by means of a command-and-control (C&C) channel. A common approach is to use an Internet Relay Chat (IRC) channel for C&C: all the bots and a botmaster join a channel and the botmaster uses the channel to broadcast commands, with responses being sent back via broadcast or private messages to the botmaster. The IRC protocol is designed to support large groups of users and a network of servers to provide scalability and resilience to failures, thus it forms a good fit for providing a C&C infrastructure. Because of their simple design and deployment, IRC botnets have been widely used by cybercriminals since 2001 (Kharouni, 2009). Some botnets use a more advanced structure, with bots communicating directly with each other in a peer-to-peer fashion, but recent studies show that *many existing botnets use the IRC model* because of its simple-yet-effective structure (Kharouni, 2009; Zhuge et al., 2007). In this research we focus on the IRC botnets.

Much research has been devoted to the detection of IRC botnets (Binkley and Singh, 2006; Ramachandran et al., 2006; Karasaridis et al., 2007; Collins et al., 2007; Villamarín-Salomón and Brustoloni, 2009; Zilong et al., 2010). However, most effective detection techniques are complex and have potential to generate false positives. This means that organizations with a large security budget are able to find potential bot infections and disable, investigate, and disinfect affected machines. Organizations with less developed IT practices, as well as home users, however, remain vulnerable to bot infections and provide a fertile ground for botnets, allowing them to remain strong.

We propose a technique that follows a service model. It leverages the efforts of one organization to capture and instantiate bot instances to provide low-cost detection of bots in other networks. We develop BotMosaic—a watermark that, when inserted into the communication between the captured bots and an IRC server, creates a pattern that is observable at other sites hosting botnets. The pattern can be recognized simply by observing the timings of the packets in a given flow, thus the detection can be carried out at a large scale by border routers. By inserting an artificial pattern, we can ensure that false-positive rates are very low, enabling automated actions to disconnect infected bots. Since only packet timings are used, BotMosaic works even when the botnet uses encrypted connections to the IRC server.

The watermark will be visible on all connections between the bots and the IRC server. It will likewise appear in the connection from the botmaster to the IRC server. Botmasters typically use stepping stones (Zhang and Paxson, 2000) to hide their true location. The watermark can be used to detect such stepping stones and aid in botmaster traceback.

A novel and unique feature of our watermark is that it is *collaborative*: the watermark is inserted simultaneously into the flows of all captured bots. This is in contrast to past watermarks that affect a single flow at a time (Wang and Reeves, 2003; Wang et al., 2005; Pyun et al., 2007; Wang et al., 2007; Yu et al., 2007; Houmansadr et al., 2009b; Ramsbrock et al., 2008). The collaborative feature amplifies the effect of the watermark and is necessary to create a timing pattern that is recognizable among the noise generated by traffic from other bots. In fact, *none* of the previous flow watermarks can be used for botnet detection application targeted in this paper (Wang and Reeves, 2003; Wang et al., 2005; Pyun et al., 2007; Wang et al., 2007; Yu et al., 2007; Houmansadr et al., 2009b; Ramsbrock et al., 2008): a bot connection watermarked using a non-collaborative scheme gets destroyed once it is mixed with flows from other bots in the C&C channel, whereas the collaborative watermarking of BotMosaic is able to persist in the mixed C&C traffic of a botnet.

In summary, BotMosaic has the following unique features as compared to previous approaches: 1) BotMosaic is implemented by one organization, and can be used as a *low-cost* service by other organizations, i.e., *client*s. A client organization only needs to deploy the low-cost watermark detectors of BotMosaic on their border routers. This is in contrast to other approaches that suggest each organization to deploy its own, resource-intensive botnet detection mechanism. 2) A client organization can use BotMosaic to detect various instances of bots simultaneously, without the need to modify its BotMosaic detectors for different botnets. The BotMosaic watermarkers use different watermark signals for different instances of botnets. 3) Each client organization can detect not only the bot infected machines, but also the botmasters and stepping stones hosts residing *inside* their networks.

We analyze our scheme using simulations and experiments on Planet-Lab (Bavier et al., 2004). We find that we can achieve a high rate of detec-tion with few false positives using a watermark applied to captured/imitated bots that comprise a small fraction of the botnet, with a detection time of about a minute.

The rest of the paper is organized as follows: Section 2 describes previous work on IRC botnet detection and reviews past work on network flow watermarking. Section 3 describes the overall detection framework used by BotMosaic. Section 4 describes the detailed structure of the BotMosaic collaborative watermark. Simulations and implementation results are presented in Section 5. Section 6 offers a brief discussion of some additional issues, and Section 7 concludes the paper.

## 2. Related work and motivation

The primary goal of the BotMosaic is to detect bot-infected machines inside a network of interest, e.g., an ISP. The literature on this can be divided into *host-based* and *network-based* approaches. Host-based approaches analyze the information on hosts of the network; this is not easy to deploy on all hosts, especially in organizations where computers are not centrally managed. BotMosaic falls in the network-based category.

Network-based detection mechanisms aim to detect bot infected machines by analyzing the network traffic information. These mechanisms mainly are classified into two categories: *traffic signature* schemes and *traffic classification* schemes. The traffic signature approaches use the captured bots to develop signatures for each botnet instance; they have widely been used for IRC botnet detection (Binkley and Singh, 2006; Karasaridis et al., 2007; Goebel and Holz, 2007). As an example, Blinkley and Singh (Binkley and Singh, 2006) combine IRC statistics and some TCP metrics to generate signatures that can be used to detect the infected machines. Traffic classification approaches are based on gathering network traces and clustering them in order to detect botnets based on their behavioral difference with the normal traffic (Ramachandran et al., 2006; Villamarín-Salomón and Brustoloni, 2009; Collins et al., 2007). As an example, Villamar et al. use Bayesian methods to isolate centralized botnets, based on the similarity of their DNS traffic with those of some known DNS botnet traces (Villamarín-Salomón and Brustoloni, 2009).

In this paper we consider a third approach for performing network-based bot detection. BotMosaic uses *network flow watermarking* to mark the botnet traffic, resulting in low-cost mechanisms for the detection of bots and botmasters. Network flow watermarking is a technique that actively perturbs the traffic patterns of a network flow to insert a watermark inside them that can later be detected. Flow watermarking has been used to detect stepping stones, as well as to compromise anonymous communication (Wang and Reeves, 2003; Wang et al., 2005; Pyun et al., 2007; Wang et al., 2007; Yu et al., 2007; Houmansadr et al., 2009b). Existing techniques, however, cannot be applied to the problem of bot/botmaster traceback for two reasons. First, they are designed to work on long-lived flows; typically, hundreds of packets are necessary to detect the presence of a watermark. Botnet communication, however, tends to be short-lived, with only a few packets sent from each bot. Furthermore, a watermark that is applied to a single bot-to-botmaster/botnet communication will be overwhelmed by traffic from other bots that will be aggregated along the same stepping stone connection. Although some of the existing watermarks are designed to resist a reasonable level of chaff, they do so by increasing the length of the watermark and thus cannot be used for botnet traceback in practice.

More recently, Ramsbrock et al. (Ramsbrock et al., 2008) designed a watermark specifically targeted to the task of botmaster traceback. Their watermark works by adding extra whitespace at the end of IRC messages sent by the bots. They also adjusted the timings of packets in order to improve detection ability. Though an important first step, the whitespace watermarking approach has several serious limitations. Whitespace watermarking only works well in the presence of low rates of chaff—less than 0.5 packets/second—whereas even in a small-size botnet, an aggregate response from all the bots would create a significantly higher chaff rate. Whitespace watermarking is also fragile to repacketization or retransmission of packets, as such events can cause it to lose timing synchronization. Finally, whitespace watermarking relies on modifying the *contents* of the messages sent by the bots, which can be be difficult if encrypted connections are used.

*Network flow watermarking.* Recently, researches have proposed to use network flow watermarks in different applications. Wang et al. were the first to borrow the watermarking idea from multimedia literature to do active traffic analysis (Wang and Reeves, 2003). They use QIM watermarks over interpacket delays (IPD) of the network flows, providing a more efficient scheme for the detection of stepping stone attacks compared to similar passive detection schemes (Wang et al., 2002).

To make the detection scheme robust to packet-level modification several watermarking schemes suggest an interval-based approach (Pyun et al., 2007; Yu et al., 2007; Wang et al., 2007). In particular, Pyun et al. proposes an interval-based watermark for detection of stepping stones which is robust to repacketization (Pyun et al., 2007). A similar interval-based scheme is proposed in Yu et al. (2007) that utilizes packet rates for watermarking. Wang et al. propose another interval-based flow watermark to compromise anonymity in low-latency anonymous networks (Wang et al., 2007). Kiyavash et al. in (Kiyavash et al., 2008) introduce a multi-flow attack that is able to compromise the interval-based watermarks of (Pyun et al., 2007; Wang et al., 2007; Pyun et al., 2007) . Houmansadr et al. uses a non-blind approach to improve the invisiblity and robustness of the watermarks (Houmansadr et al., 2009b). Ramsbrock et al. use flow watermarking for the real-time traceback of botmaster in IRC based botnets, by inserting additional whitespaces at the end of IRC messages (Ramsbrock et al., 2008). Houmansadr et al. propose SWIRL, a scalable and robust watermark that takes a flow-dependent watermarking approach (Houmansadr and Borisov, 2011). As mentioned before, the existing watermarking schemes can not be used for the the problem of bot/botmaster traceback for two reasons. First, they are designed to work on long-lived flows; typically, hundreds of packets are necessary to detect the presence of a watermark. Botnet communication, however, tends to be short-lived, with only a few packets sent from each bot. Second, a watermark that is applied to a single bot-to-botmaster/botnet communication will be overwhelmed by traffic from other bots that will be aggregated along the same stepping stone connection. Although some of the existing watermarks are designed to resist a reasonable level of chaff, they do so by increasing the length of the watermark and thus cannot be used for botnet traceback in practice.

## 3. BotMosaic detection framework

In this section we describe the features of IRC botnets exploited by BotMosaic and its deployment scenarios.

### 3.1. IRC botnets

Internet Relay Chat (IRC) is a network protocol designed for Internet text messaging or synchronous conferencing (Oikarinen and Reed, 1993; Kalt, 2000). In order to use an IRC network, clients *join* an IRC *channel* created by an IRC server, providing a *nickname* (the server may also require client authentication). The clients then send broadcast messages to the channel, or private messages to specific nicknames inside that channel.

During the last decade, IRC channels have been used as a common way to construct the C&C channel of botnets; the rational behind this traditional decision by cybercriminals is the low weight of IRC client software, the simplicity of the IRC protocol, and the existence of many public IRC servers over the Internet that can be used by the botnets (Kharouni, 2009; Zhuge et al., 2007). Some examples of IRC-based botnets are SdBot, Virut, SpyBot, and RBot (Zhuge et al., 2007). The infected bot hosts act as IRC clients and join a specific channel used by the botnet. Some botnets use fixed channels, while other change them dynamically in order to avoid detection and shutdown. The bot then communicates with the botmaster and other bots using the IRC channel.

The botmaster sends commands to the bots by sending a broadcast message to the channel, or by sending private messages to individual bots. For example, a botmaster might send messages such as "send me recorded passwords" or "start DDoS on target $X$." The bot will send responses back either as private messages (for sensitive data, such as credit card information) or public broadcasts (for, e.g., status updates). To avoid detection, bots may encrypt the contents of the messages and/or use an encrypted connection to the IRC server.

### 3.2. BotMosaic Architecture

In BotMosaic, a *service provider* inserts watermarks using captured bots that are then detected by a number of *client organization*s (shortly, *client*s). The service provider performs the majority of the work, whereas the clients run low-cost detection. The service provider may either charge clients to use the watermark, by selling a subscription to the watermark secret keys, or provide it as a public service.

**Captured Bots:** BotMosaic relies on a number of bot instances that are controlled by the service provider. These bots will be used to insert the watermark pattern into the IRC channel. To capture these bots, the service provider may deploy a honeynet (Spitzner, 2003) or manually infect a number of (possibly virtual) machines with the bot.

**Watermarker:** The watermarker mediates the traffic between the captured bots and the outside world to insert the watermark. In particular, it delays network packets to create a particular timing pattern. The watermarker will delay packets from *all* captured bots simultaneously to ensure that the watermark signal is strong enough to be detected. Note that the watermark is content-agnostic; BotMosaic will work even if the traffic between the bots and the IRC server is encrypted.

**Detector:** A detector, run by the client, watches network traffic for the watermark pattern. Typically it would be deployed at or near border routers, to examine all the traffic entering and leaving the client's organization. It only needs to examine packet timings and headers (the latter to group packets into flows) to detect the mark; importantly, it does not need to perform deep packet inspection. The detection algorithms can thus be run efficiently on high-speed network links.

### 3.3. Value Proposition

The BotMosaic clients will benefit from the scheme in several important ways, creating incentives for deployment of the scheme by ISPs and enterprises.

#### 3.3.1. Detecting Bots

When the watermarker inserts a mark onto broadcast messages from the captured bots, this watermark will be observed on the traffic from the IRC server to other bots. A client running a BotMosaic detector can, therefore, detect bots hosted in its own network by monitoring incoming traffic for watermarks, such as in network $A$ in Fig. 1. The detector is much lower cost than other methods of botnet detection, and the watermark parameters can be configured to ensure an acceptably low rate of false positives.

#### 3.3.2. Detecting Stepping Stones

A botmaster will typically use a number of *stepping stones* to connect to the IRC channel, in order to disguise his or her identity (Zhang and Paxson, 2000). The stepping stones will carry the watermark as well; a client can discover a stepping stone hosted within its network by observing the watermark on an *outgoing* flow, as in network $B$ in Fig. 1.

A stepping stone is usually a compromised computer, thus detecting stepping stones is valuable to the client. This information can also be used to help locate the botmaster: if the first stepping stone used by the botmaster is in an organization running BotMosaic detection, the IP address of the botmaster will be revealed. Note that this remains true even if all of the other stepping stones are on networks not covered by BotMosaic.

Our approach is a variant on other techniques that use watermarks for stepping stone detection (Pyun et al., 2007; Wang et al., 2007; Houmansadr et al., 2009b). However, previous work required an organization to insert watermark on all inbound traffic. BotMosaic, in contrast, does not require clients to modify or delay traffic flows and thus will not interfere with the level of service provided to legitimate users. It can be deployed on a mirrored port, whereas watermark insertion must be performed inline, creating a potential point of failure.

### 3.3.3. Detecting the Botmaster

Finally, a client hosting the botmaster will be able to observe the watermark on its inbound connection, as in network $C$ in Fig. 1. The botmaster will receive the watermark both on broadcast messages and on private responses from bots (as long as the private responses are sent by all bots in response to a command). Thus, one way to distinguish the botmaster from ordinary bots is that, in some instances, the botmaster will be able to observe the watermark even though other bots did not.

## 4. BotMosaic watermarking scheme

In this section, we describe the watermarking scheme that we devise to be used for the BotMosaic botnet traceback system. The watermark is novel in being *collaborative*: the BotMosaic service provider uses *multiple* captured bots for watermark insertion, which makes the scheme specialized for the problem of botnet watermarking. Multiple captured bots allow us to spread the watermark power over a larger amount of traffic, compensating for the small amount of traffic each individual bot sends to the botmaster/botnet. BotMosaic adopts an interval-based design used in other watermarks (Pyun et al., 2007; Wang et al., 2007; Yu et al., 2007) to provide robustness to packet losses, delays, and reordering as well as chaff introduced by the traffic of the other bots in the botnet.

### 4.1. Watermark insertion

Fig. 2 shows the structure of the BotMosaic service provider. The BotMosaic service provider starts $R$ captured bots, joining and communicating with the botnet through the IRC C&C channel. Such connections are intercepted by a watermarker, as described later. Using virtual machines (Provos, 2004), $R$ can be made reasonably large while using modest amounts of resources. Let $B$ be the number of *active* real bots connecting through the same C&C channel. Increasing the ratio $R/B$ improves watermark detection efficiency, as will be shown in the following sections.

Fig. 3 illustrates the collaborative watermark insertion on the communication of captured bots. Each of the flows contain a share of the watermark so that the mixture of the packets of these flows, combined in the botnet C&C channel, generates the watermark pattern, whereas any single captured bot flow is insufficient to detect the watermark.

The watermarker inserts a watermark sequence of length $l$ into the captured bots flows. The watermarker divides the time axis into $2l$ non-overlapping intervals with equal length $T$. The intervals are labeled as $HI_1, \ldots, HI_l$ and $LO_1, \ldots, LO_l$ using a random assignment. The interval mappings form the secret watermark key that is necessary to detect the watermark; the service provider can therefore sell watermark key subscriptions to clients.

The basic idea of the watermark is to send more packets in HI interval compared to its corresponding LO interval in the mixture of all captured flows. For each HI-LO pair, the watermarker assigns the timing of $R$ captured flows so that in the mixture of all $R$ captured flows the number of packets appearing in the $HI_i$ interval of that is larger than the number of packets in the corresponding $LO_i$ interval by some threshold $\eta$ (see Fig. 3). In other words, we should have:

$$\sum_{f=1}^{R} N_f(HI_i) - \sum_{f=1}^{R} N_f(LO_i) \geq \eta \qquad i = 1, \ldots, l \qquad (1)$$

where $N_f(\cdot)$ gives the number of packets showing up in the given interval of $f^{th}$ captured flow

For the manipulation of the captured flows the watermarker first determines the total number of packets in each interval of the aggregated watermarked flow and then assigns the number of packets to each captured flow, namely *watermark shares*. Based on IRC standards (Kalt, 2000) the rate of the flows from IRC client to the server should not exceed 0.5 packets per second (exceeding this threshold causes the client to be penalized by extra delays on its packets). So, in an interval with length $T$ seconds we expect to have at most $\frac{T \cdot R}{2}$ packets accumulated over all $R$ captured flows. For each $i$, we randomly select the cumulative packet number for $HI_i$ interval to be:

$$N(HI_i) \epsilon \left[ \frac{T \cdot R}{4}, \frac{T \cdot R}{2} \right] \qquad (2)$$

We then assign the total number of packets in the $LO_i$ interval to be:

$$N(LO_i) = N(HI_i) - \eta - \psi \qquad (3)$$

where $\eta$ is the *detection threshold* and $\psi$ is the *confidence threshold*. Doing so for all of the HI-LO pairs we find the cumulative number of packets $N(j)$ for any interval $j$. We then randomly distribute the $N(j)$ packets within the $j^{th}$ interval of all $R$ captured flows so that the overall rate of each flow does not exceed the 0.5 packets per second constraint mentioned above. For each captured flow, the watermarker buffers a number of packets before starting the watermark insertion, to make sure it has an adequate number of packets for its manipulations.

## 4.2. Detection Scheme

The watermark detectors deployed on the border routers of the BotMosaic clients monitor network traffic to detect the watermark patters inserted by the BotMosaic service provider (see Section 3)). The detectors are provided with watermark key(s) by the BotMosaic service provider:

$$Key = (T, \{\forall i = 1, \ldots, l : HI_i, LO_i\}) \qquad (4)$$

As mentioned before, the flows watermarked by the BotMosaic service provider get mixed with other flows, resulting in a single mixed (and usually, encrypted) flow. We assume that the target mixed-encrypted flow contains packets from $R$ captured bots and $B$ real bots. A watermark detector breaks up a candidate flow into intervals, and then computes $N(HI_i)$ and $N(LO_i)$. The detector then calculates the following:

$$\Delta(i) = N(HI_i) - N(LO_i), \qquad \text{for } i = 1, \ldots, l \qquad (5)$$

The detector uses two thresholds to decide whether a watermark is present. First, if $\Delta(i) > \eta$ ($\eta$ is the detection threshold in (3)), the detector calls the $i^{th}$ pair in the sequence *detected*. Note that the confidence threshold $\psi$ used during the watermark insertion ensures that natural variations in numbers of packets do not destroy the watermark.

Finally, the detector declares the candidate flow to be watermarked if the total number of detected pairs $n_c$ (out of $l$) is greater than or equal to some threshold $\theta$, *hamming threshold*. It is easy to see that by increasing $\eta$ and $\theta$, we can decrease the number of false positives at the cost of creating more false negatives. We will discuss parameter choices in Section 5. Due to the delays applied to the mixed watermarked flows passing through the network, the detectors need to perform *synchronization*. This is done using sliding windows, as will be discussed in section 6.1.

## 5. Simulations and Experiments

In the simulations and experiments of this section we only consider the detection of BotMosaic watermarks being inserted into the traffic towards the botmaster. The detection of watermarks on traffic to bots is similar; however, botmaster detection is more difficult as the botmaster traffic is relayed through a number of stepping stones, resulting in more delays affecting the watermark pattern.

### 5.1. Simulations

We simulated BotMosaic in Matlab to evaluate its performance. We used the traces of botnets for *SpyBot* and *SdBot* botnets that were also used in BotMiner research Gu et al. (2008). The SdBot trace belongs to a botnet with a botmaster and four real bots. Since we needed a larger botnet to evaluate our scheme, we extended the trace to have 100 real bots. Based on analyzing the trace, bots listen on the IRC channel for the commands, and upon receiving a command they respond to it appropriately. To extend the botnet trace from the existing 4-bot trace to a 100-bot trace, we sent a response to the channel on behalf of the newly added bots after a random delay whenever a command was issued and the existing bots responded to it.

To simulate the watermarking scheme, we added watermarked flows generated by the rogue bots to the trace for different settings of watermark parameters. For each run of the simulations, we generated a new extended trace, selecting a different part from the real trace randomly and extending the trace for 100 bots as discussed above. We insert the watermark into the trace and calculate the number of pairs that are detected by the detection scheme (true positive pairs); we also run detection on the unwatermarked version of the trace to and count the number of detected pairs (false positive pairs). This lets us estimate the error rates for a given threshold $\theta$; namely, how many watermarked flows would *not* be detected (false-negative rate) and how many non-watermarked flows would be misdetected (false-positive rate). We then adjust $\theta$ so that false-positive and false-negative error rates are equal; the resulting rate is called the *crossover error rate* (COER) and we call the corresponding threshold $\hat{\theta}$.

We also add delay and jitter to the botnet traffic, based on measurements we have performed on PlanetLab Bavier et al. (2004). In Section 6.1 we will discuss how to synchronize our detector with the watermarker. However, non-uniform delay for different bots, as well as network jitter, will decrease the accuracy of our detection and thus we include it in our simulation. Each experiment is run 100 times (each time with the same watermark parameters but different watermark key and different bot traces) to get the mean and variance of true-positive and false-positive parameters. Using these statistics we estimate the COER for each experiment by approximating the false error rate distributions with normal distributions. The Kolmogorov-Smirnov (K-S) test indicates that the true-positive and false-positive parameters are fitted to normal distributions with average significance levels of 0.0121 and 0.045, respectively. The average K-S distances from a normal distribution are 0.0808 and 0.0680, respectively. In our experiments, we set the number of (active) real bots to be $B = 100$, and vary the number of rogue bots, $R$. Fig. 4a illustrates the estimated COER versus the watermark length, for different values of the parameter $T$. The $R/B$ value is fixed to 10%. In all of the simulations, the detection threshold $\eta$ and the confidence threshold $\psi$ are set to 1.

We also estimated the $COER$ for different values of $R/B$ ratio. Fig. 4b shows the $COER$ for different watermark lengths having different ratios of $R/B$. As expected, increasing $R/B$ improves the $COER$ at the expense of requiring more resources to run a larger number of instances.

Table 1 shows the results of the experiment for two different settings of the watermark parameters (each experiment is run 500 times). For the interval length of $T = 500\,\text{ms}$ and using 64 pairs and for $R/B = 10\%$, a watermark can be inserted into a 64 second connection with the botmaster, and the resulting COER is on the order of $10^{-8}$, which is very promising. Increasing the $T$ parameter improves the COER, at the expense of needing more time for the botmaster to be online. We find that $\hat{\theta}$ is approximately $l/2$.

We also performed similar experiments over SpyBot traces from Gu et al. (2008), leading to similar results. Table 1 also shows the detection results for two sample sets of watermarking parameters for the SpyBot traces. As can be seen the watermark can be detected in as few as 64 seconds with a COER of about $10^{-8}$. Similar to SdBot simulations, we can trade elapsed time for COER, using different values of the watermarking parameters.

Table 1: Two sample runs of the detection scheme over SdBot and SpyBot traces, for $R/B = 10\%$ and a watermark sequence of length 64 (averaged over 100 random runs).

| Botnet type | T | True pairs | False pairs | $\hat{\theta}$ | COER | Elapsed time (s) |
|---|---|---|---|---|---|---|
| SdBot | 250 | 43.9 | 22.4 | 33 | $2.8 * 10^{-3}$ | 32 |
| | 2000 | 51.3 | 12.9 | 32 | $3.52 * 10^{-13}$ | 256 |
| SpyBot | 500 | 48.0 | 16.2 | 32 | $2.32 * 10^{-8}$ | 64 |
| | 2000 | 50.3 | 13.5 | 32 | $7.55 * 10^{-11}$ | 256 |

## 5.2. Implementation

We tested BotMosaic on PlanetLab by creating synthetic bots that use an IRC channel to communicate with the botmaster. The captured bots are implemented over physically separate hosts in PlanetLab. Watermark proxies are installed in front the captured bot hosts, and are controlled by a controller to insert the watermark. We route all the bot traffic through the watermark proxy. Watermark proxies are responsible for watermarking the bot traffic on all the captured bot machines, so that the accumulated traffic makes the final watermarked traffic. By using a proxy, we avoid having to reverse engineer and modify the bot code to insert watermarks.

We implemented the BotMosaic watermarking scheme over the Planet-Lab infrastructure using randomly selected nodes as different entities in the experiment. Fig. 5 shows the structure of our experiment A botmaster is controlling botnet through the IRC C&C. To hinder detection, the botmaster relays his traffic to the IRC server through 5 stepping stone nodes located in geographically different locations, and also encrypts the connections between stepping stones.

There are 100 real bots ($B = 100$) connected to the IRC C&C channel, listening for the commands from the botmaster and sending appropriate responses to the channel. The real bots are chosen randomly, and are located in geographically diverse locations. We set up $R = 10$ captured bots to send watermarked flows to the C&C channel ($R/B = 10\%$). The captured bots are also chosen randomly and are located in different places. A controller node commands the captured bots to join the C&C channel. Once all the captured bots have joined the channel, the controller commands all of them to start sending packets on the C&C channel containing corresponding shares of the watermark.

Again, we only provide the results for the watermark inserted on the botnet traffic to the botmaster, i.e., through PRIVMSG to the botmaster; the detection performance is the same for the watermark inserted into the traffic directed to the bots.

We set up several watermark detectors across the network to look for the inserted watermark in network flows. The detector deployment is described in Section 3. We set up 5 detectors on the way to the botmaster to check the true detection rate, and also 3 detectors on the paths not leading to the botmaster to evaluate the false detection rate.

Fig. 6 shows detection results for different detectors. We set the $T$ parameter of the watermarking system to be $T = 500\,\text{ms}$ and use watermarks with sequence length $l$ equal to 32, 64, and 128. According to the simulations in the previous section, we set the hamming threshold $\theta$ to be $l/2$ in each case. Results are normalized by the sequence length.

As the results show, detectors on the path from IRC server to the botmaster ($MN\#1$ to $MN\#5$) are able to detect the watermark from the mixed-encrypted flows, as soon as only 32 seconds. On the other hand, detectors placed on the paths not containing the botmaster watermark do not detect the watermark on the innocent flows.

## 6. Discussion

We briefly discuss several other issues regarding the BotMosaic scheme.

### 6.1. Detector synchronization

Watermark detectors need to synchronize the received watermarked flow with the watermark sequence, i.e., minimize the offset between intervals of watermarked flow and those of the watermark, in order to successfully detect the watermark. To find the right offset of the watermarked flow, we run the watermark detection scheme over the received flow applying different offset values from 0 to $T$ in $T/100$ steps, and select the offset maximizing the number of detected pairs as the right offset for that flow. Our experiments show that running the synchronization mechanism over an non-watermarked flow, the number of detected pairs remains below the hamming threshold $\theta$ for different offset values. The use of this synchronization mechanism makes detection scheme tolerate different network delays (though variable network jitter still impacts the detection accuracy).

Table 2: Resources required for BotMosaic.

| $l$ | Processing Time ($\mu$sec/flow) | Total Memory per flow (KB) | Total Memory (MB) |
|-----|-----|-----|-----|
| 32 | 28.0 | 0.16 | 3.46 |
| 64 | 49.3 | 0.27 | 6.06 |
| 128 | 86.9 | 0.48 | 10.49 |

## 6.2. Resources

In order to study the processing and memory costs of BotMosaic detection scheme we ran the watermark detector over a 21 GB network trace gathered from the routers of an anonymous US university. The utilized trace contains 21 744 concurrent flows, with a total of 2.1 GB of timing information. Since the detection scheme should be implemented over border routers, this volume of traffic is representative of a highly loaded border router.

The experiment is done using a Unix system with a 1.6 GHz Intel CPU and 2 GB of memory. Table 2 illustrates the result of the experiment over the university traces. Even for a watermark of length 128, which would provide very low error rates, the total memory needed for watermark detection is about 11 MB, and the processing time for each flow is as only 87 $\mu$s. The time and processing resources are even smaller for shorter watermarks. Thus we expect that it may be possible to deploy BotMosaic even in high-performance routers used by large ISPs, to provide a better vantage point for bot detection and botmaster traceback.

## 6.3. Watermark evasion

As with any flow watermarking scheme, an attacker who wishes to foil watermark detection can do so by inserting large delays and other modifications to the flow structure. Therefore, an adversary with full control over an IRC server can render a botnet immune to BotMosaic. There are other evasion avenues available to the botnet designer, including the use of a peer-to-peer structure or covert communication (Nagaraja et al., 2011) for C&C. Our goal, however, is to capture a class of existing bots that, despite using simple and well-understood C&C techniques, still comprises a large fraction of current botnets seen in the wild Kharouni (2009); Zhuge et al. (2007). Forcing botnet operators to use more advanced C&C mechanisms imposes new costs and affects the profitability of the entire criminal enterprise.

### 6.4. Other issues

Traditional flow watermarking schemes consider issues like chaff, repacketization, and packet addition/removal on the performance of watermark detection. BotMosaic is designed to be robust to interfering traffic from non-captured bots and the corresponding mechanisms likewise address the above issues.

Interval-based watermark schemes are subject to a multi-flow attack discovered by Kiyavash et al. (Kiyavash et al., 2008). This might allow a non-subscriber to recover the secret watermark key; if this is a concern, we can use the mechanisms proposed by Houmansadr et al. (Houmansadr et al., 2009a), for example, by changing the watermark key (the set of HI and LO intervals) over the time.

Finally, it might be the case that the botmaster puts limitations on the number of packets each bot can send. In this case, watermarking will still be feasible by using more captured bots to collaborate in the generation of BotMosaic watermarks.

## 7. Conclusion

We have presented a new botnet traceback scheme, BotMosaic, that detects bot infected machines and helps to track down the botmasters controlling the centralized botnets. BotMosaic uses a service-based approach where detector clients perform fast and low-cost watermark detection, which is much cheaper and easier to deploy than existing signature- and classification-based detectors. We presented a new collaborative flow watermarking structure, making it suitable for the botnet detection problem. We showed through experiments that our watermark can be quite effective when 5%-10% of the bots are captured/imitated by a service provider, and that our detection is simple enough to be able to handle large volumes of traffic. Any individual organization deploying the low-cost BotMosaic detectors can realize bot defense benefits, providing an incremental path to widespread deployment of the BotMosaic architecture and potential detection of botmasters.

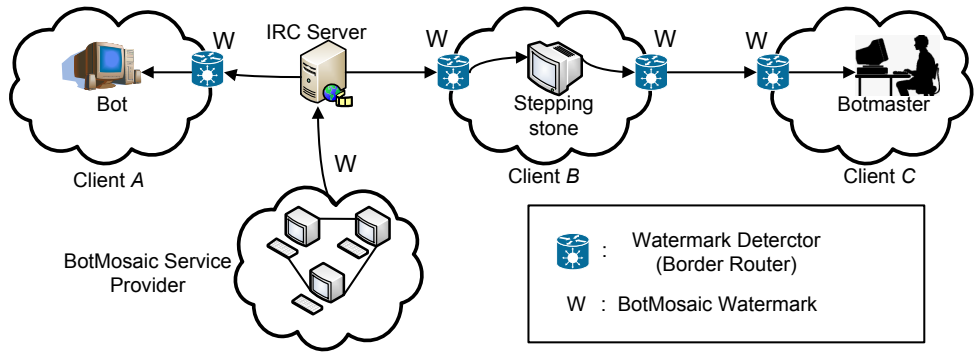

Figure 1: Topology of the BotMosaic traceback system.

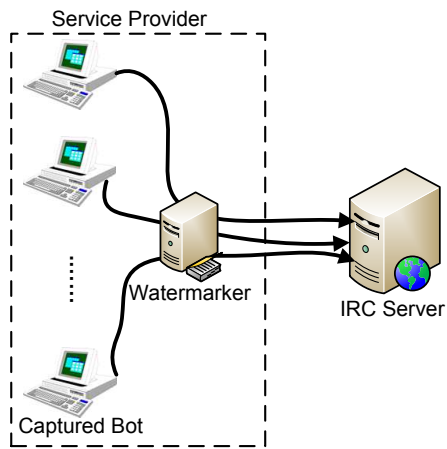

Figure 2: BotMosaic service provider structure.

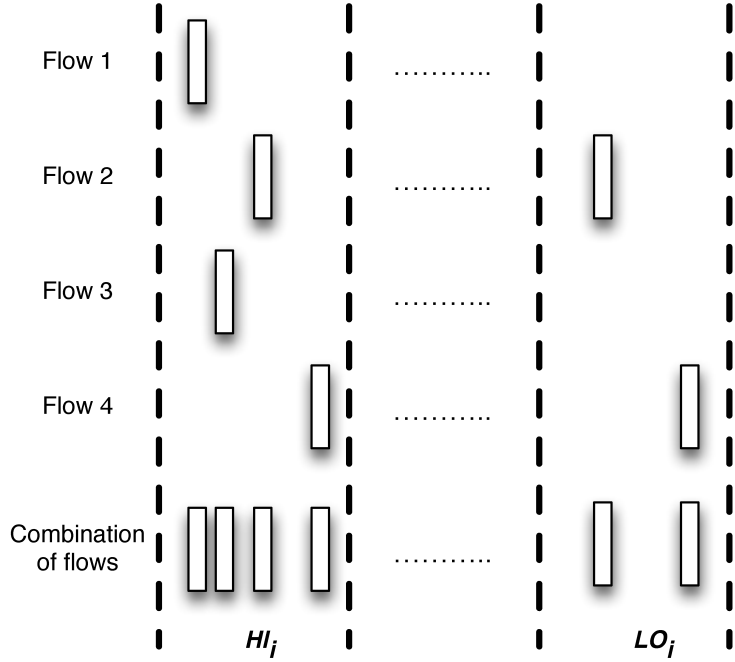

Figure 3: Collaborative watermark insertion using multiple flows.

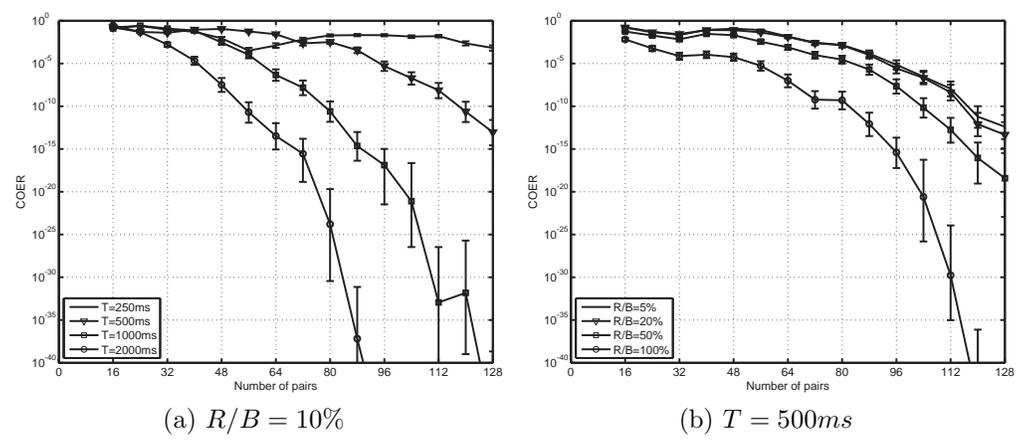

(a) $R/B = 10\%$           (b) $T = 500ms$

Figure 4: COER error of the detection scheme over SdBot botnet traces along with 95% confidence intervals.

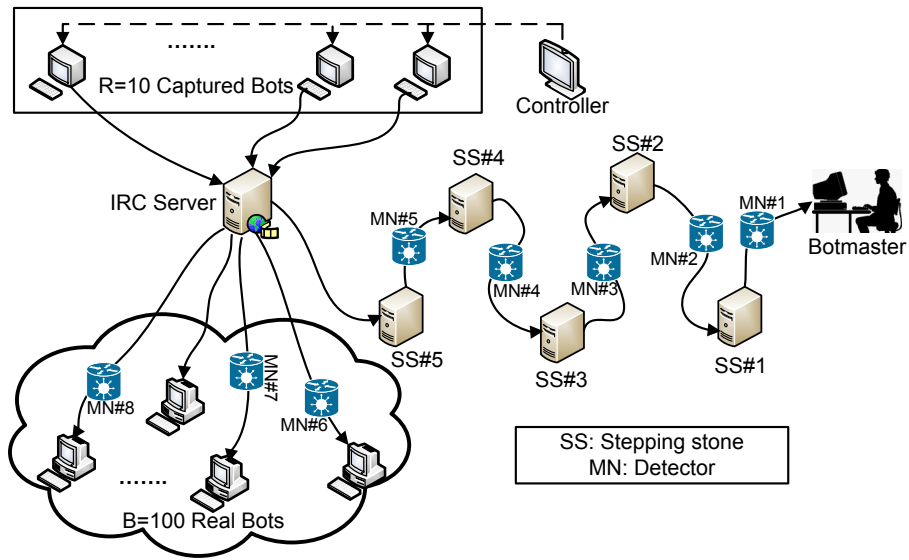

Figure 5: Testbed structure of BotMosaic implementation over PlanetLab.

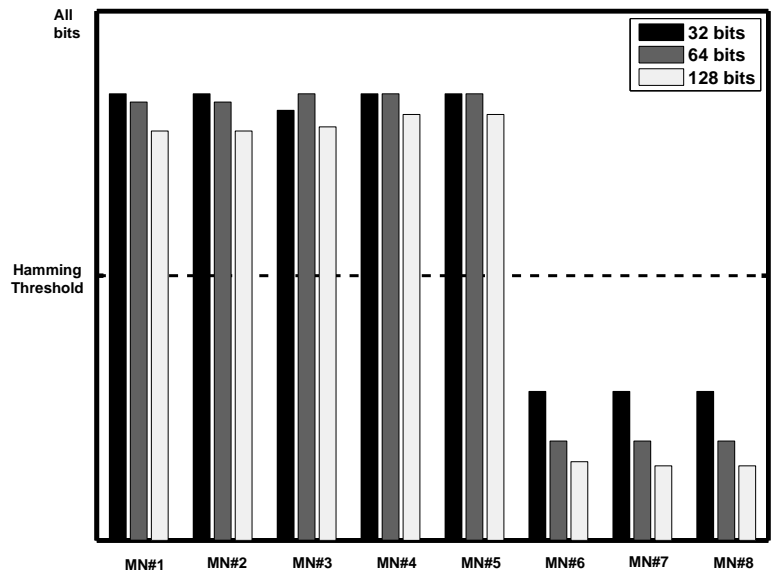

Figure 6: Watermark pairs detected by different detectors across the testbed (normalized by total number of pairs).

24